\documentclass[runningheads]{llncs}
\usepackage{graphicx}
\begin{document}
\title{Automated ischemic stroke lesion segmentation from 3D MRI \protect\\ ISLES 2022 challenge report}
%
%
\author{Md Mahfuzur Rahman Siddiquee \and
Dong Yang \and
Yufan He \and
Daguang Xu \and
Andriy Myronenko
}

\authorrunning{M. Rahman Siddiquee et al.}
%
\institute{NVIDIA, Santa Clara, CA \\
\email{\{mdmahfuzurr,dongy,yufanh,daguangx,amyronenko\}@nvidia.com}}
\maketitle             

\begin{abstract}
Ischemic Stroke Lesion Segmentation challenge (ISLES 2022) offers a platform for researchers to compare their solutions to 3D segmentation of ischemic stroke regions from 3D MRIs. In this work, we describe our solution  to ISLES 2022 segmentation task. We re-sample all images to a common resolution,  use two input MRI modalities (DWI and ADC) and train SegResNet semantic segmentation network from MONAI. The final submission is an ensemble of 15 models (from 3 runs of 5-fold cross-validation).  Our solution (team name NVAUTO) achieves the top place in terms of Dice metric (0.824), and overall rank 2 (based on the combined metric ranking\footnote{https://isles22.grand-challenge.org/}). It is implemented  with Auto3DSeg\footnote{https://monai.io/apps/auto3dseg}.

\keywords{ISLES22  \and  MICCAI22 \and segmentation challenge \and MONAI \and Auto3Dseg \and SegResNet \and 3D MRI.}
\end{abstract}

\section{Introduction}

\begin{figure}[!ht]
    \centering
    \includegraphics[width=0.32\textwidth]{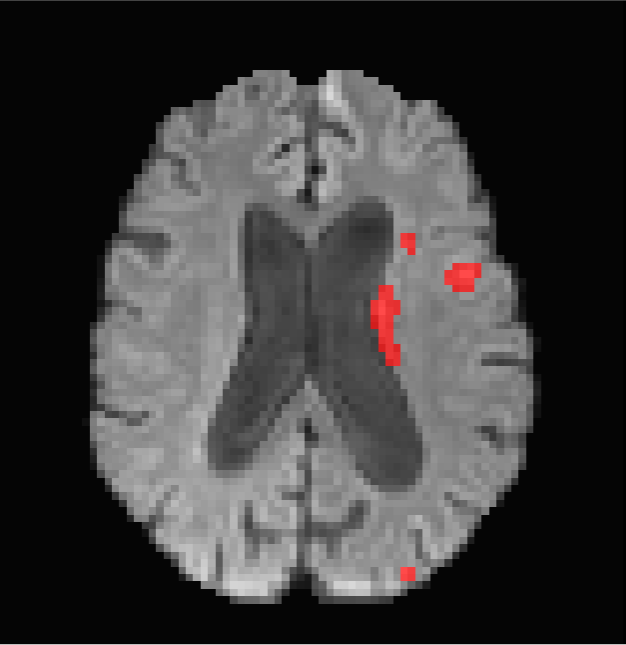}
    \includegraphics[width=0.32\textwidth]{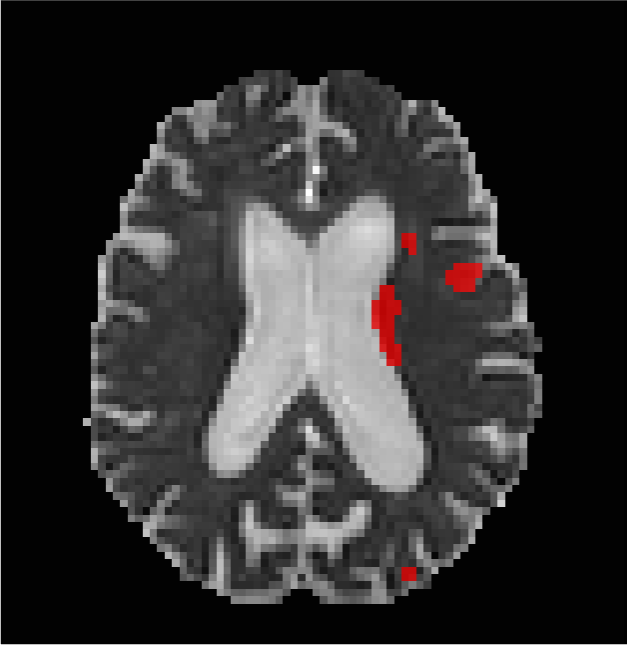}
    \includegraphics[width=0.32\textwidth]{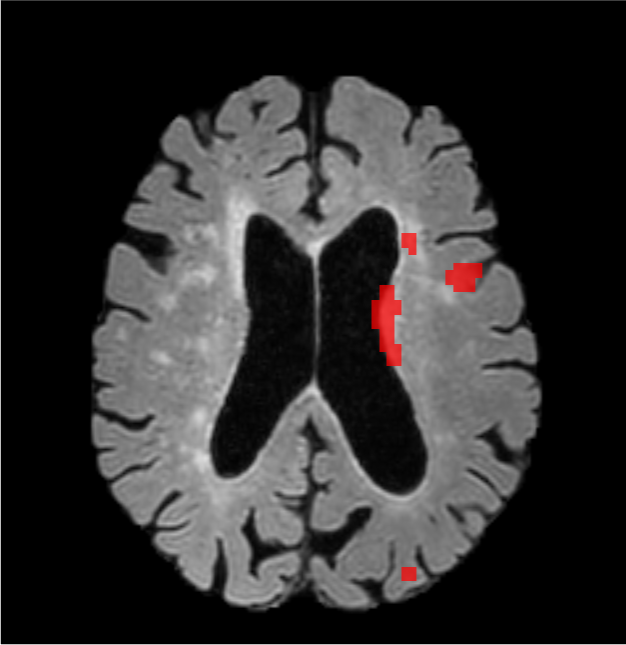}
    \caption{An example of 3D MRIs (DWI, ADC and FLAIR) images showing ischemic stroke region (in red).}
    \label{fig:example1}
\end{figure}

Segmentation of ischemic stroke region is necessary for treatment planning and evaluation of patients' disease outcomes.  Ischemic Stroke Lesion Segmentation challenge (ISLES 2022) aims to benchmark infarct segmentation in acute and sub-acute stroke using multimodal 3D MRI modalities~\cite{hernandez2022isles}. The ISLES22 dataset consists of 400 cases (250 labeled labeled cases were provided for training, and 150 reserved for testing). Each case includes three 3D MRI modalities ( DWI, ADC and FLAIR), and the task is to segment a single class (stroke region). The images were skull stripped and intensity normalized by the organizers. The ground truth labels were provided in the reference space of DWI and ADC image modalities. The FLAIR image modality is at a higher resolution, and also  slightly misaligned from the reference frame (in this work we decided not use the FLAIR modality, and  use only DWI and ADC images). An example case with 3D MRIs and the corresponding ground-truth overlays is shown in Figure~\ref{fig:example1}.

\section{Method}

\begin{figure}[t]
    \centering
    \includegraphics[width=0.8\textwidth]{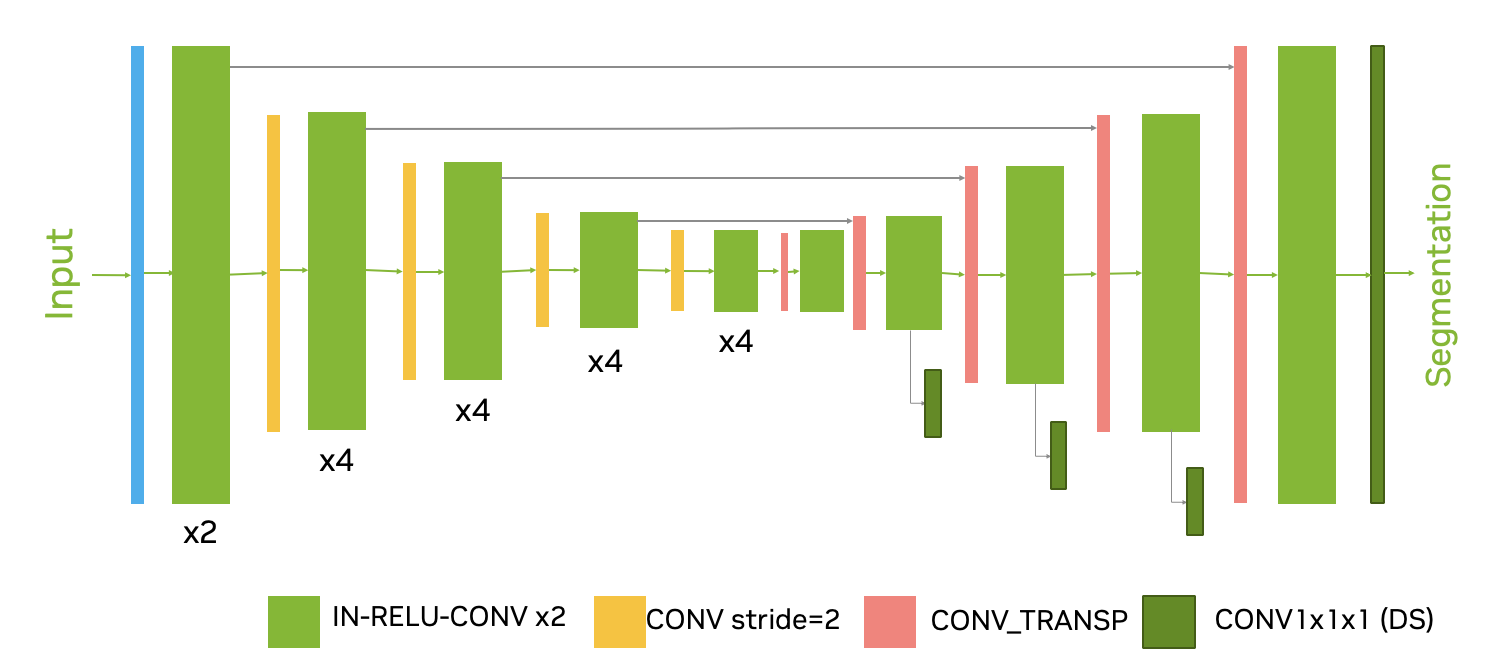}

    \caption{SegResNet network configuration. The network uses repeated ResNet blocks with Instance normalization, and deep supervision in the decoder branch}
    \label{fig:net}
\end{figure}

We implemented our approach with MONAI\footnote{https://github.com/Project-MONAI/MONAI}~\cite{monai}, we used Auto3Dseg\footnote{https://monai.io/apps/auto3dseg} system to automate most parameter choices. For the main network architecture we used SegResNet\footnote{https://docs.monai.io/en/stable/networks.html\#segresnet}, which is an encode-decoder based semantic segmentation network based on~\cite{myronenko20183d}, with deep supervision (see Figure~\ref{fig:net}).


The encoder part uses ResNet~\cite{he2016identity} blocks with instance normalization. We have used 5 stages of down-sampling, each stage has 2, 4, 4, 4, and 4 convolutional blocks, respectively. Each block's output is followed by an additive identity skip connection. We follow a common CNN approach to downsize image dimensions by 2 progressively and simultaneously increase feature size by 2.  All convolutions are 3x3x3 with an initial number of filters equal to 32. The encoder is trained with $192\times192\times128$ input region. The decoder structure is similar to the encoder one, but with a single block per each spatial level. 
 The end of the decoder has the same spatial size as the original image, and the number of features equal to the initial input feature size, followed by a 1x1x1 convolution and a softmax. For the deep supervision of 3 sub-levels of the decoder branch features, we add extra projections heads (1x1x1 convolutions) into the same number of output classes. These additional network outputs are used to compute  losses at smaller image scales. 

\subsubsection{Dataset} We use the ISLES dataset~\cite{hernandez2022isles}  for training the model.  We randomly split the entire dataset into 5 folds and trained a model for each fold. 

\subsubsection{Data preparation} We use only DWI and ADC image modalities, re-sample them to a common 1x1x1mm resolution, and concatenate into a two channel 3D input. 

\subsubsection{Data normalization} We normalize each MRI image to a zero mean and unit std.

\subsubsection{Cropping} We crop a random patch of 192x192x128 voxels, which includes most of the foreground content. We use the batch size of 1.

\subsubsection{Augmentations} We use random a) flips (all axes)
b) random rotation and scaling,  c) random smoothing, noise, intensity scale/shift. 

\subsubsection{Loss} We use the combined Dice + Focal loss. The same loss is summed over all deep-supervision sublevels:

\begin{equation}
Loss= \sum_{i=0}^{4} \frac{1}{2^{i}} Loss(pred,target^{\downarrow}) 
\end{equation}
where the weight $\frac{1}{2^{i}}$ is smaller for each sublevel (smaller image size) $i$. The target labels are downsized (if necessary) to match the corresponding output size using nearest neighbor interpolation. 

\subsubsection{Optimization} We use the AdamW optimizer with an initial learning rate of $2e^{-4}$ and decrease it to zero at the end of the final epoch using the Cosine annealing scheduler. All the models were trained for 1000 epochs with deep supervision. We use batch size of 1 per GPU, and train on 8 GPUs 16Gb NVIDIA V100 DGX machine (which is equivalent to batch size of 8). We use weight decay regularization of $1e^{-5}$.

\subsubsection{Pretraining}  We pretrain the network on Brats21 dataset using the same network configuration~\cite{brats21}. In our experiments, the pretraining slightly increases the cross-validation dice accuracy (0.5-1\%).

\section{Results}

Based on our data splits, a single run 5-folds cross-validation results are shown in Table~\ref{tab:result}.  On average, we achieve $0.8086$ cross-validation performance in terms of  dice metric (see~Table~\ref{tab:result}).

\begin{table}[h!]
    \centering
    \begin{tabular}{| c | c | c | c | c | c |}
        \hline
        {\textbf{Fold 1}} & {\textbf{Fold 2}} & {\textbf{Fold 3}} & {\textbf{Fold 4}} & {\textbf{Fold 5}} & {\textbf{Average}} \\
        \hline
         0.8232 & 0.7948 & 0.797 & 0.8118 & 0.8163 & 0.8086 \\
        \hline
    \end{tabular}
    \caption{Average DICE among classes using 5-fold cross-validation.}
    \label{tab:result}
\end{table}

For the final submission we use a mean ensemble of 15 models total (3 fully trained runs, using best checkpoints). Table~\ref{tab:result2} shows the final ranking and scores on the hidden test sets (provided by the organizers) for the top 3 places. Our solution (NVAUTO) team achieves the best dice accuracy, but ranks 2nd overall based on all 4 metrics used in this challenge (Dice, F1-score, Average volume difference, lesion count difference). 
 
\begin{table}[h!]
    \centering
    \begin{tabular}{| l | l | c | c | c | c |}
        \hline
        \textbf{Rank} & \textbf{Team} & \textbf{Dice} & \textbf{F1-score} & \textbf{Volume diff} &  \textbf{Lesion count diff}  \\
        \hline
        1 & SEALS~\cite{team_seals22} & 0.821 & 0.857 & 1.634 & 1.0\\
        \hline
        2 & NVAUTO (ours) & 0.824 & 0.8 & 1.634 & 2.0 \\
        \hline
        3 & Factorizer~\cite{team_factorizer22} & 0.812 & 0.8 & 1.936 & 2.0\\
        \hline
    \end{tabular}
    \caption{Top 3 teams of the ISLES22 challenge with the corresponding 4 metrics used for the final ranking. }
    \label{tab:result2}
\end{table}

\bibliographystyle{splncs04}
\bibliography{paper}

\begin{thebibliography}{1}
\providecommand{\url}[1]{\texttt{#1}}
\providecommand{\urlprefix}{URL }
\providecommand{\doi}[1]{https://doi.org/#1}

\bibitem{monai}
Project-monai/monai, \url{https://doi.org/10.5281/zenodo.5083813}

\bibitem{team_factorizer22}
Ashtari, P.: Factorizer: marrying matrix factorization with deep learning for
  stroke lesion segmentation. In: {ISLES22 report, MICCAI2022}. KU Leuven,
  Belgium (2022)

\bibitem{brats21}
Baid, U., Ghodasara, S., Bilello, M., Mohan, S., Calabrese, E., Colak, E.,
  Farahani, K., Kalpathy{-}Cramer, J., Kitamura, F.C., Pati, S., Prevedello,
  L.M., Rudie, J.D., Sako, C., Shinohara, R.T., Bergquist, T., Chai, R., Eddy,
  J., Elliott, J., Reade, W., Schaffter, T., Yu, T., Zheng, J., Annotators, B.,
  Davatzikos, C., Mongan, J., Hess, C., Cha, S., Villanueva{-}Meyer, J.E.,
  Freymann, J.B., Kirby, J.S., Wiestler, B., Crivellaro, P., Colen, R.R.,
  Kotrotsou, A., Marcus, D.S., Milchenko, M., Nazeri, A., Fathallah{-}Shaykh,
  H.M., Wiest, R., Jakab, A., Weber, M., Mahajan, A., Menze, B.H., Flanders,
  A.E., Bakas, S.: The {RSNA-ASNR-MICCAI} brats 2021 benchmark on brain tumor
  segmentation and radiogenomic classification. CoRR  \textbf{abs/2107.02314}
  (2021), \url{https://arxiv.org/abs/2107.02314}

\bibitem{team_seals22}
Gao, S., Zhang, S., Liang, K., Liu, Y., Zhang, Z., Li, Z., Yin, Z.: Solution of
  team {SEALS for ISLES22}. In: {ISLES22 report, MICCAI2022}. Deepwise AI Lab
  and Beijing University of Telecommunication, Beijing, China (2022)

\bibitem{he2016identity}
He, K., Zhang, X., Ren, S., Sun, J.: Identity mappings in deep residual
  networks. In: European conference on computer vision. pp. 630--645. Springer
  (2016)

\bibitem{hernandez2022isles}
Hernandez~Petzsche, M.R., de~la Rosa, E., Hanning, U., Wiest, R.,
  Valenzuela~Pinilla, W.E., Reyes, M., Ines~Meyer, M., Liew, S.L., Kofler, F.,
  Ezhov, I., et~al.: Isles 2022: A multi-center magnetic resonance imaging
  stroke lesion segmentation dataset. arXiv e-prints pp. arXiv--2206 (2022)

\bibitem{myronenko20183d}
Myronenko, A.: {3D MRI} brain tumor segmentation using autoencoder
  regularization. In: International MICCAI Brainlesion Workshop. pp. 311--320.
  Springer (2018)

\end{thebibliography}
\end{document}